\documentclass{PoS}

\usepackage{slashed}
\usepackage[numbers]{natbib}
\usepackage{rotating}
\usepackage{graphicx}
\newcommand{\met}{\slashed{E}_T}
\newcommand{\ET}{\slashed{E}_T}

\title{EW scale DM models with dark gauge symmetries}

\ShortTitle{EW scale DM models with dark gauge symmetries}

\author{\speaker{Pyungwon Ko}\thanks{}\\ 
        School of Physics, KIAS, Seoul 02455, Korea \\
        E-mail: \email{pko@kias.re.kr}}

\abstract{
In this talk, I describe a class of electroweak (EW) scale dark matter (DM) models where its
stability or longevity are the results of underlying dark gauge symmetries: stable due to 
unbroken local dark gauge symmetry or topology, or long-lived due to the accidental global
symmetry of dark gauge theories. Compared with the usual phenomenological dark matter 
models (including DM EFT or simplified DM models),  DM models with local dark gauge 
symmetries include dark gauge bosons, dark Higgs bosons and sometimes excited dark matter.  
And dynamics among these fields are  completely fixed by local gauge principle. 
The idea of singlet portals including the Higgs portal  can thermalize these hidden sector 
dark matter very efficiently, so that these DM could be easily thermal DM.  I also discuss 
the limitation of the usual DM effective field theory or simplified DM models without the full 
SM gauge symmetry,  and emphasize the importance of the full SM gauge symmetry and 
renormalizability especially for collider searches for DM. 
}

\FullConference{The 11th International Workshop Dark Side of the Universe 2015\\
		14-18 December 2015\\
		Kyoto, Japan}

\begin{document}



\section{INTRODUCTION}

The standard model (SM) has been tested  from atomic scale up to $\sim O(1)$ TeV scale 
by many experiments, and has been extremely successful. 
However, there are some observational facts which call for new physics beyond the SM 
(BSM): (i) baryon number asymmetry of the universe (BAU), (ii) neutrino masses and mixings, 
(iii) nonbaryonic dark matter (DM) and (iv) inflation in the early universe.

In this talk, I will concentrate on the issue of DM, assuming that BAU and neutrino masses 
and mixings are accommodated by the standard seesaw mechanism by introducing heavy 
right-handed (RH) neutrinos. 
For the inflation, I assume that the Higgs inflation is a kind of minimal setup, and 
I show that the dark Higgs from hidden sectors can modify the standard Higgs inflation in a
such a  way that a larger tensor-to-scalar ratio $r \sim O(0.01 - 0.1)$ independent of 
precise values of the top quark and the SM Higgs boson mass \cite{Ko:2014eia}. 

First of all, I discuss the basic assumption for DM models, emphasizing the role of dark gauge
symmetry, renormalizability, unitarity and limitation of DM effective field theory (EFT). 
Then I give specific examples where (i) DM is absolutely stable due to unbroken dark gauge 
symmetry or topological reason, and (ii) DM is long-lived due to accidental global symmetry
of underlying dark gauge symmetry. One of the common features of these models is the 
existence of a new neutral scalar boson from dark sector, which I will call dark Higgs boson.
I show that dark Higgs boson can play a new key role in HIggs inflation, EW vacuum stability, 
light mediator generating self-interaction of DM, and explaining the galactic center 
$\gamma$-ray execss. 
This talk is based on a series of my works ~\cite{Hur:2007uz,Ko:2008ug,Hur:2011sv,Baek:2011aa,Baek:2012uj,Baek:2012se,Baek:2013qwa,Chpoi:2013wga,Baek:2013dwa,Ko:2014nha,
Ko:2014bka,Ko:2014gha,Ko:2014eia,Baek:2014jga,Ko:2014loa,Baek:2014kna,Ko:2015eaa,Ko:2015ioa,Baek:2015lna} with various collaborators. 

\section{Basic assumptions for DM models}

\subsection{Relevant questions for DM} 
So far the existence of DM was confirmed only through the astrophysical and cosmological 
observations where only gravity play an important role.  
described by quantum field theory (QFT), 
We have to seek for the answers to the following questions for better understanding of DM:
\begin{itemize}
\item How many species of DM are there in the universe ?
\item What are their masses and spins ?
\item Are they absolutely stable or very long-lived ?
\item How do they interact among themselves and with the SM particles ?
\item Where do their masses come from ? 
\end{itemize}
In order to answer (some of) these questions, we have to observe its signals from colliders 
and/or various (in)direct detection experiments.


The most unique and important property of DM (at least, to my mind) is that DM particle should be 
absolutely stable or long-lived enough, similarly to the case of electron and proton in the SM.  
Let us recall that electron stability is accounted for by electric charge conservation (which is exact), 
and this implies that there should be massless photon, associated with unbroken $U(1)_{\rm em}$ gauge symmetry.   On the other hand, the longevity of proton is ascribed to the baryon number which is an 
accidental global  symmetry of the SM, broken only by dim-6 operators.  
We would like to have DM models where DM is absolutely stable or long-lived enough by similar reasons 
to electron and proton.  And this special property of DM has to be realized in the fundamental Lagrangian 
for DM in a proper way in QFT, similarly to QED and the SM.  Local dark gauge symmetry will play important
roles, by gauranteeing the stability/longevity of DM, as well as determine dynamics in a complete and 
mathematically consistent manner.

\subsection{Hidden sector DM and local dark gauge symmetry}

Any new physics models at the electroweak scale are strongly constrained
by electroweak precision test and CKM phenomenology, if new particles  
feel SM gauge interactions.  The simplest way to evade these two strong constraints is to assume 
a weak scale hidden sector which is made of particles neutral under the SM gauge interaction. 
A hidden sector particle could be a good candidate  for nonbaryonic dark matter of the universe, 
if it is absolutely stable or long lived.  Note that hidden sectors are very generic in many BSMs, including
SUSY models. The hidden sector matters may have their own gauge interactions, which we call 
dark gauge interaction associated with local dark gauge symmetry $G_{\rm hidden}$. 
They can be easily thermalized if there are suitable messengers  between the SM 
and the hidden sectors.  We also assume all the singlet operators such as Higgs portal 
or $U(1)$ gauge kinetic mixing play the role of messengers. 

Another motivation for local dark gauge symmetry $G_{\rm hidden}$ in the hidden 
sector is to stabilize the weak scale DM particle by dark charge conservation laws, 
in the same way electron is absolutely stable because it is the lightest charged 
particle and electric charge is absolutely conserved.  %

Finally note that all the observed particles in Nature feel some gauge interactions in addition 
to gravity.  Therefore it looks very natural to assume that dark matter of the universe (at least some
of the DM species) also feels some (new) gauge force, in addition to gravity.

\subsection{EFT vs. Renormalizable theories} 

Effective field theory (EFT) approaches are often adopted for DM physics.
For example, let us consider a singlet fermion DM model in EFT: 
\begin{equation}
{\cal L}_{\rm fermion DM}  =   \overline{\psi} \left[  i \not \partial - m_\psi \right] \psi 
- \frac{\lambda_{H\psi}}{\Lambda} H^\dagger H \overline{\psi} \psi 
\end{equation}
with {\it ad hoc} discrete $Z_2$ symmetry under $\psi \rightarrow - \psi$.
However this could be erroneous for a number of reasons.  

Let us consider one of its UV completions \cite{Baek:2011aa}: 
\begin{eqnarray}
{\cal L}_{\rm DM} & = & {1 \over 2} (\partial_\mu S \partial^\mu S - m_S^2 S^2) 
-\mu_S^3 S - {\mu_S^\prime \over 3} S^3  -  {\lambda_S \over 4} S^4  \nonumber \\
& + &  \overline{\psi} ( i \not \partial - m_\psi ) \psi 
- \lambda S \overline{\psi} \psi  
 -  \mu_{HS} S H^\dagger H -{\lambda_{HS} \over 2} S^2 H^\dag H . 
\label{eq:Lag2}
\end{eqnarray}
We have introduced a singlet scalar $S$ in order to make the model (1) renormalizable. 
There will be two scalar bosons $H_1$ and $H_2$ (mixtures of $H$ and $S$) in our model,   
and the additional scalar $S$ makes the DM phenomenology completely different from 
those from Eq. (1).  This is also true for vector DM models~\cite{Baek:2012se,Ko:2014gha}.

For example, the direct detection experiments such as XENON100 and LUX exclude
thermal DM within the EFT model (1), but this is not true within the UV completion 
(2), because of generic cancellation mechanism in the direct detection due to 
a generic destructive interference between $H_1$ and $H_2$ contributions for fermion or vector DM
~\cite{Baek:2011aa,Baek:2012se}.  
Also the direct detection cross section in the UV completion is related with 
that in the EFT  by~\cite{Baek:2014jga}
\begin{equation}
\sigma_{\rm SI}^{\rm ren} = \sigma_{\rm SI}^{\rm EFT} ~
 \left( 1 - \frac{m_{125}^2}{m_1^2} \right)^2~ \cos^4 \alpha \ ,
\end{equation}
which includes the cancellation mechanism and corrects the results reported by ATLAS 
and CMS (see Fig. ~1).
Here $m_1$ is the mass of the singlet-like scalar boson and $m_{125}$ is the Higgs 
mass found at the LHC.  Note that the EFT result is recovered when $\alpha \rightarrow 0$ and 
$m_1 \rightarrow \infty$.  

\begin{figure}[h]
\centering
\includegraphics[width=0.45\textwidth]{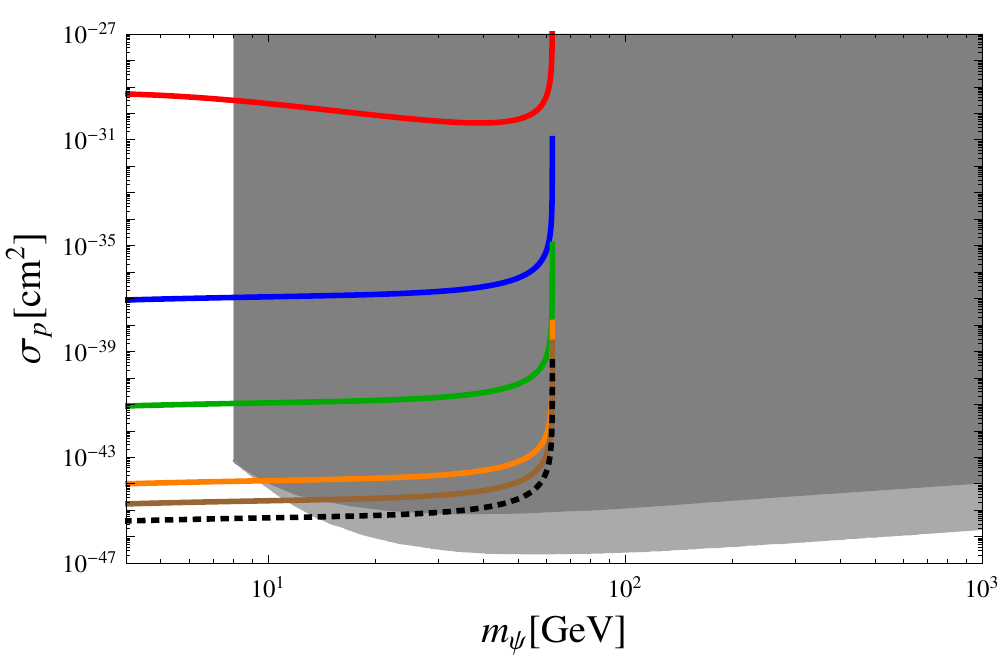}
\includegraphics[width=0.45\textwidth]{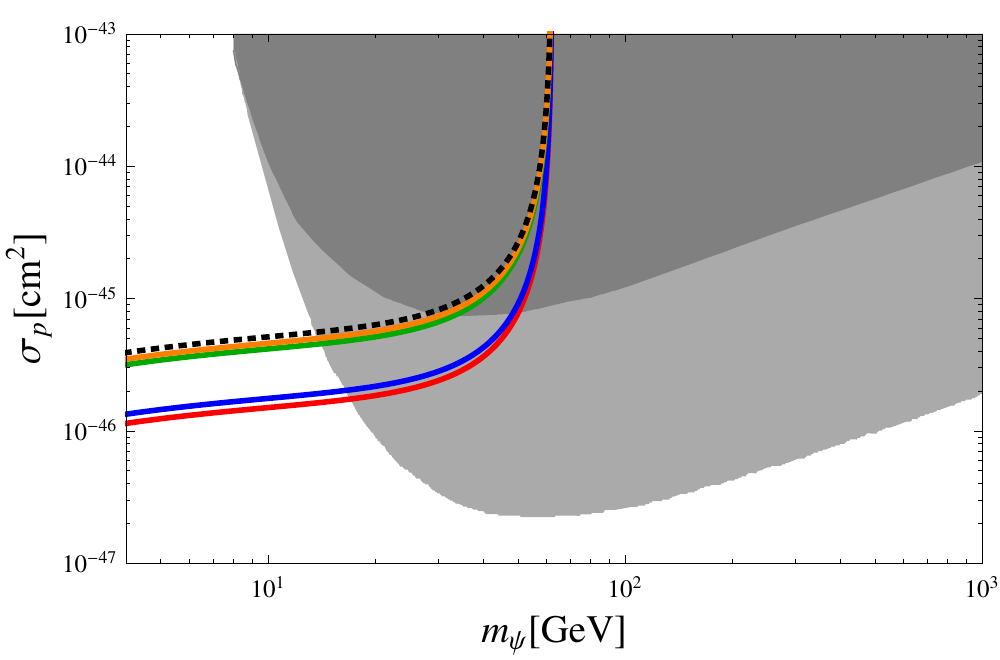}
\includegraphics[width=0.45\textwidth]{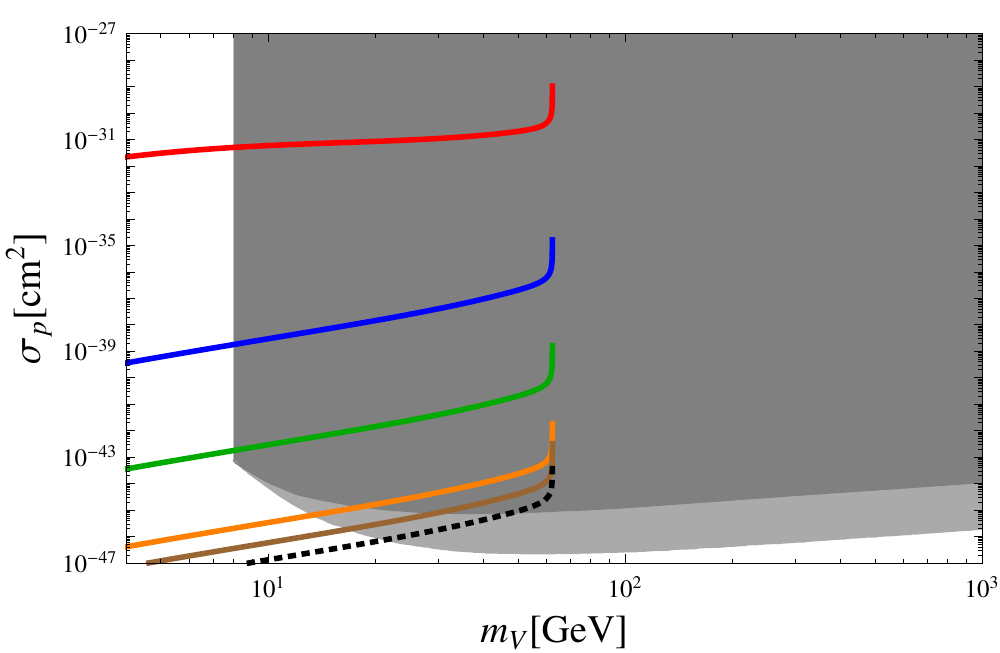}
\includegraphics[width=0.45\textwidth]{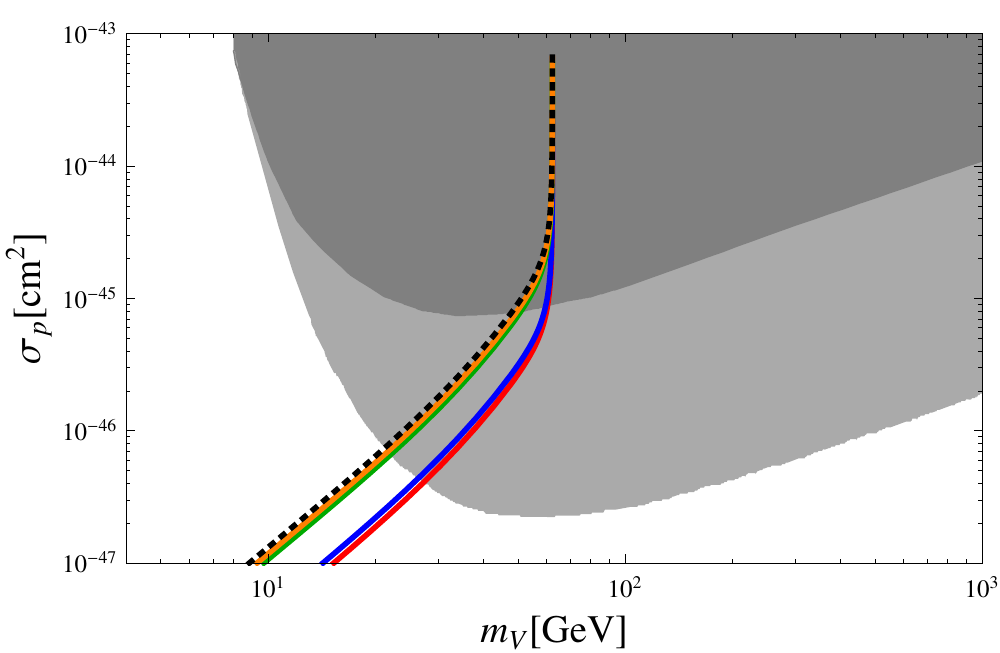}
\caption{\label{fig:sp-SFDM} $\sigma_p^{\rm SI}$ as a function of the mass of dark matter 
for SFDM (top) and VDM (bottom) for a mixing angle $\alpha=0.2$.
Left panel: $m_2 = 10^{-2}, 1, 10, 50, 70$ GeV for solid lines from top to bottom.
Right panel: $m_2 = 100, 200, 500, 1000$ GeV for dashed lines from bottom to top.
The black dotted line is EFT predictions presented by ATLAS and CMS 
\cite{Aad:2014iia,Chatrchyan:2014tja}.  
Dark-gray and gray region are the exclusion regions 
of LUX and projected  XENON1T (gray).} 
\end{figure}


\subsection{Dark Higgs mechanism for the vector DM and $\gamma$-ray excess from the GC}

One can also consider Higgs portal DM both in EFT and in a unitary and renormalizable model 
\cite{Baek:2012se}, where dark Higgs is naturally introduced. It can be shown that one can accommodate
the GeV-scale $\gamma$-ray excess from the GC very easily in terms of VDM annihilating into a pair of
dark Higgs \cite{Ko:2014gha}:$VV\rightarrow H_2 H_2$, followed by $H_2$ decays into the SM particles. 
This new mechanism is in fact very generic in hidden sector DM models with local dark gauge symmetries 
\cite{Ko:2014loa}.   More details are discussed in a talk by Yong Tang at this meeting~\cite{Ko:2015eaa}.
Finally a recent study shows that the best fit to the $\gamma$-ray sepctra is obtained if $M_{\rm DM} \simeq 
95$GeV, $M_{H_2} \simeq 86.7$ GeV and  $\langle \sigma v \rangle \simeq 4 \times 10^{-26}$cm$^3 /s$
with a p-value $=0.40$~\cite{Ko:2015ioa}. Such a dark Higgs is very difficult to study at colliders, and
indirect signatures of DM could be a nice complementary.

The Higgs portal VDM model is usually described by 
\begin{equation}
{\cal L}_{\rm VMD} = - \frac{1}{4} V_{\mu\nu} V^{\mu\nu} + \frac{1}{2} m_V^2 V_\mu V^\mu  
 - \frac{\lambda_{HV}}{2} V_\mu V^\mu |H|^2 - \frac{\lambda_V}{4!} V^4
\end{equation}
with an ad hoc $Z_2$ symmetry, $V_\mu \rightarrow -V_\mu$.   Although all the operators are either dim-2 
or dim-4, this Lagrangian breaks gauge invariance, and is neither unitary nor renormalizable. 

One can consider the renormalizable Higgs portal vector DM model by introducing a dark Higgs 
$\Phi$ that generate nonzero mass for VDM by the usual Higgs mechanism: 
\begin{equation}
{\cal L}_{VDM}  =  - \frac{1}{4} X_{\mu\nu} X^{\mu\nu} +  
(D_\mu \Phi)^\dagger (D^\mu \Phi)   - \lambda_\Phi \left( | \Phi |^2 - \frac{v_\Phi^2}{2} \right)^2
 - \lambda_{\Phi H} \left( | \Phi |^2 - \frac{v_\Phi^2}{2}\right) \left( | H |^2  - \frac{v_H^2}{2}
\right) \ ,
\label{eq:full_theory}
\end{equation}
Then the dark Higgs from $\Phi$ mixes with the SM Higgs boson in a similar manner as in 
SFDM.   And there is a generic cancellation mechanism in the direct detection cross section. 
Therefore one can  have a wider range of VDM mass compatible with both thermal relic density 
and direct detection cross  section  (see Ref.~\cite{Baek:2012se} for more details). 
In particular the dark Higgs can play an important role  in DM phenomenology. 

Another important observable is the Higgs invisible decay width. 
The invisible Higgs decay width in the EFT VDM model is given by 
\begin{equation}
( \Gamma_h^{\rm inv} )_{\rm EFT} 
= \frac{\lambda_{VH}^2}{128 \pi} \frac{v_H^2 m_h^3}{m_V^4} \times
 \left( 1 - \frac{4 m_V^2}{m_h^2} + 12 \frac{m_V^4}{m_h^4} \right) 
 \left( 1 - \frac{4 m_V^2}{m_h^2} \right)^{1/2} .
\end{equation} 
Note that the invisible decay rate in the EFT becomes arbitrarily large as 
$m_V \rightarrow 0$,  which is not physical. 
Let us compare this with the invisible  Higgs decay in the renormalizable and unitary 
Higgs portal VDM model, whcih is given by 
\begin{equation}
\Gamma_i^{\rm inv} = \frac{g_X^2}{32 \pi} \frac{m_i^3}{m_V^2} \left( 1 - \frac{4 m_V^2}{m_i^2} 
+ 12 \frac{m_V^4}{m_i^4} \right) \left( 1 - \frac{4 m_V^2}{m_i^2} \right)^{1/2} .
\end{equation}
where $m_V$ is the mass of VDM.   In this case $m_V = g_X v_\Phi$ so that the invisible
decay width does not blow up when $m_V \rightarrow 0$, unlike the EFT VDM case.
This is another example demonstrating the limitation of the EFT calculation.   
%

\begin{figure}
\centering
\includegraphics[scale=0.5]{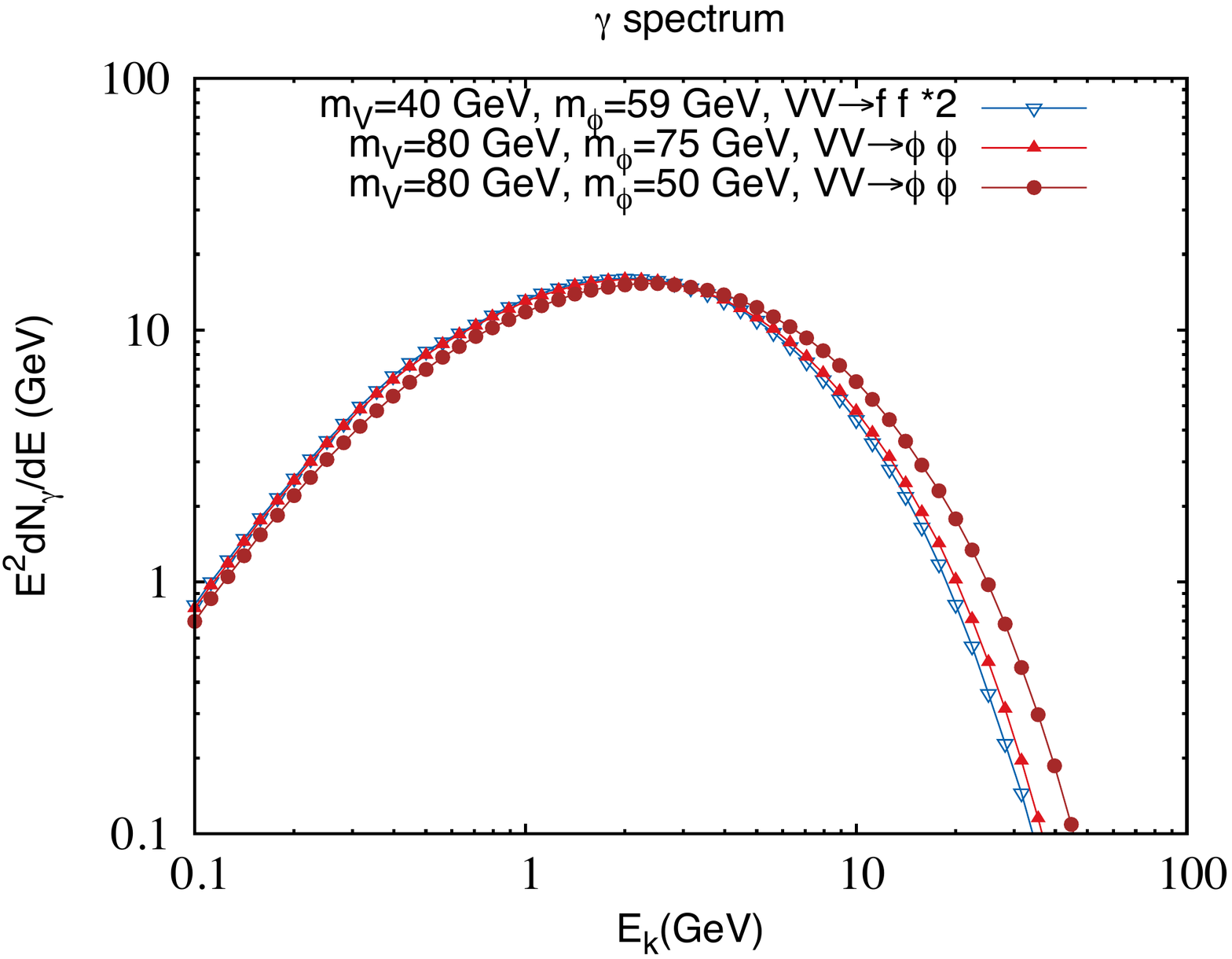}
\caption{Illustration of $\gamma$ spectra from different channels. The first two cases give almost the same 
\label{fig:gammaspectra}}
\end{figure}

Having the dark Higgs can be very important in DM phenomenology. 
Let me demonstrate it in the context of the GeV scale $\gamma$-ray excess from the galactic center (GC). 
In the Higgs portal VDM with dark Higgs, one can have a new channel for $\gamma$-rays: namely, 
$V V \rightarrow H_2 H_2$ followed by $H_2 \rightarrow b\bar{b}, \tau \bar{\tau}$ through a small mixing
between the SM Higgs and the dark Higgs.  As long as $V$ is slightly heavier than $H_2$ with 
$m_V \sim 80$GeV, one can reproduce the $\gamma$-ray spectrum similar to the one obtained from 
$VV\rightarrow b\bar{b}$ with $m_V \sim 40$GeV (see Fig.~\ref{fig:gammaspectra} and 
Ref.~\cite{Ko:2014gha} for more detail). Note that this mass range for VDM was not allowed within the 
EFT approach based on Eq.~(4),  where there is no room for the dark Higgs at all.  It would have been
simply impossible to accommodate the $\gamma$-ray excess from the galactic center within the 
Higgs portal VDM within EFT. Also this mechanism is generically possible in hidden sector DM models~\cite{Ko:2014loa}.

\section{Stable DM with unbroken dark gauge symmetries}

\subsection{ Local $Z_2$ scalar case}

In order to highlight the idea of local dark gauge symmetry, let us consider a scalar 
DM $S$ with Higgs  portal with discrete $Z_2$ symmetry ($S \rightarrow -S$):
\begin{equation}
{\cal L}_{\rm scalar DM}   =    \frac{1}{2} \partial_\mu S \partial^\mu S - 
\frac{1}{2} m_S^2 S^2   -   \frac{\lambda_{HS}}{2} |H|^2  S^2 - \frac{\lambda_S}{4} S^4  
\end{equation}
This model is the simplest DM model in terms of the number of new degrees of freedom 
beyond the SM, and its phenomenology has been studied comprehensively. However the 
origin and the nature of $Z_2$ symmetry has not  been specified at all in the literature. 

If this $Z_2$ symmetry is global, it could be broken by gravitation effect with $Z_2$-breaking 
dim-5 operator:
\begin{equation}
\frac{\lambda}{M_{\rm Planck}}~F_{\mu\nu} F^{\mu\nu} \ , \ \ \ \frac{\lambda}{M_{\rm Planck}}~
\bar{Q}_L H d_R  \ , \ \ \ {etc.}
\end{equation}
Then the decay rate of $S$ due to these $Z_2$-breaking dim-5 operators is given by 
\begin{equation}
\Gamma (S) \sim \frac{\lambda^2 m_S^3}{M_{Planck}^2} \sim \lambda^2 ~
\left( \frac{m_S}{100 ~{\rm GeV}} 
\right)^3  ~10^{-37}~{\rm GeV}
\end{equation}
Therefore EW scale CDM $S$ will decay very fast and cannot be a good CDM candidate,
unless the coefficient of this dim-5 operator is less than $10^{-8}$.  This is one possibility,
but another possibility is to implement the global $Z_2$ symmetry as an unbroken subgroup 
of some local dark gauge symmetry.

In fact, one can construct local $Z_2$ model, by assuming that a DM $X$ and a dark Higgs 
$\phi_X$ carry $U(1)_X$-charges equal to 1 and 2, respectively. 
The renormalizable Lagrangian of this model is given by ~\cite{Baek:2014kna}
\begin{eqnarray}
{\cal L} & =  & {\cal L}_{\rm SM}  - \frac{1}{4} \hat{X}_{\mu\nu} \hat{X}^{\mu\nu} 
-  \frac{1}{2} \sin \epsilon \hat{X}_{\mu\nu} \hat{B}^{\mu\nu} 
 +   D_\mu \phi_X D^\mu \phi_X + D_\mu X^\dagger D^\mu X  
-  \mu \left( X^2 \phi_X^\dagger +  H.c. \right) \nonumber \\
& -  & m_X^2 |X|^2  -   \lambda_X |X|^4 
- \lambda_\phi \left( | \phi_X |^2 - \frac{v_\phi^2}{2} \right)^2 
-   \lambda_{\phi X} |X|^2 |\phi_X |^2 
-  \lambda_{\phi H} | \phi_X |^2 | H |^2 -  \lambda_{HX} | X|^2  | H |^2 , 
\label{eq:model}
\end{eqnarray}
which is much more complicated than the original $Z_2$ scalar DM model, Eq. (4). 
After  $U(1)_X$ symmetry breaking by nonzero $\langle \phi_X \rangle = v_X$, 
there still remains a $Z_2$ symmetry, $X\rightarrow -X$, which guarantees the scalar
DM to be absolutely stable even if we consider higher dimensional operators. 
The $U(1)_X$ breaking also lifts the degeneracy between the real and the imaginary 
parts of $X$, $X_R$ and $X_I$ respectively.   Compared with the global $Z_2$ scalar 
DM model described by Eq. (4),   the local $Z_2$ model has three more fields: dark photon 
$Z^{'}$, dark Higgs $\phi_X$ and the excited scalar DM $X_R$, assuming $X_I$ is lighter
than $X_R$.  Then the DM phenomenology would be muvh richer than the global $Z_2$ 
scalar DM model.  For example, one can consider $X_I X_I \rightarrow \phi_X \phi_X$ followed
by $\phi_X$ decay into the SM particles through the small mixing between dark Higgs $\phi_X$ 
and the SM Higgs boson $h$, as a possible explanation of the galactic center 
$\gamma$-ray excess (see Ref.~\cite{Baek:2014kna} for more detail).

\subsection{Local $Z_3$ scalar DM model}

Let us assume the dark sector has a local $U(1)_{X}$ gauge symmetry spontaneously broken into 
local $Z_{3}$  \'{a} la Krauss and Wilczek.  
This can be achieved with two complex scalar fields $\phi_X$ and $X$   
in the dark sector with the $U(1)_{X}$ charges equal to $1$ and $1/3$, respectively
~\cite{Ko:2014nha,Ko:2014loa}.  Here $\phi_X$ is the dark Higgs that breaks $U(1)_X$ into its
$Z_3$ subgroup by nonzero VEV.  
Then one can write down renormalizable Lagrangian for the SM fields and the dark 
sector fields, $\tilde{X}_\mu, \phi_X$ and $X$: 
\begin{eqnarray}
{\cal L} & =  & {\cal L}_{{\rm SM}}-\frac{1}{4}\tilde{X}_{\mu\nu}\tilde{X}^{\mu\nu}
-\frac{1}{2}\sin\epsilon\tilde{X}_{\mu\nu}\tilde{B}^{\mu\nu}  
 +D_{\mu}\phi_{X}^{\dagger}D^{\mu}\phi_{X} 
+D_{\mu}X^{\dagger}D^{\mu}X-V (H, X, \phi_X ) \\
V & = & -\mu_{H}^{2} | H |^2 +\lambda_{H} | H^{\dagger}H |^4 -\mu_{\phi}^{2} | \phi_{X} |^2 
+\lambda_{\phi} | \phi_{X} |^4  
+\mu_{X}^{2} | X |^2 +\lambda_{X} | X |^4 
+\lambda_{\phi H} | \phi_{X} |^2 | H |^2    \nonumber \\ 
&  +& \lambda_{\phi X} | X |^2 | \phi_{X}|^2
+\lambda_{HX} |X|^2 | H |^2 + \left( \lambda_{3}X^{3}\phi_{X}^{\dagger}+H.c. \right)   \label{eq:potential}
\end{eqnarray}
where the covariant derivative associated with the gauge field $X^{\mu}$
is defined as $D_{\mu}\equiv\partial_{\mu}-i\tilde{g}_{X}Q_{X}\tilde{X}_{\mu}$. 

We are interested in the phase with the  following vacuum expectation values for the 
scalar fields in the model:
\begin{eqnarray}
\langle H\rangle=\frac{1}{\sqrt{2}}\left(\begin{array}{c}
0\\
v_{h}
\end{array}\right),\;\langle\phi_{X}\rangle=\frac{v_{\phi}}{\sqrt{2}},\;\langle X\rangle=0,\label{eq:vacuumstate}
\end{eqnarray}
where only $H$ and $\phi_{X}$ have non-zero vacuum expectation values(vev). This
vacuum will break electroweak symmetry into $U(1)_{\rm em}$, and $U(1)_{X}$ 
symmetry into local $Z_3$, which  stabilizes the scalar field $X$ and make it DM.
The discrete gauge $Z_3$ symmetry stabilizes the scalar DM even if we consider 
higher dimensional nonrenormalizable operators which are invariant under $U(1)_X$.
This is in sharp constrast with the global $Z_3$ model considered in Ref.~\cite{Belanger:2012zr}.
Also the particle contents in local and global $Z_3$ models are different so that the 
resulting DM phenomenology are distinctly different from each other, as summarized in Table 1.

\begin{figure}
\includegraphics[width=0.90\textwidth]{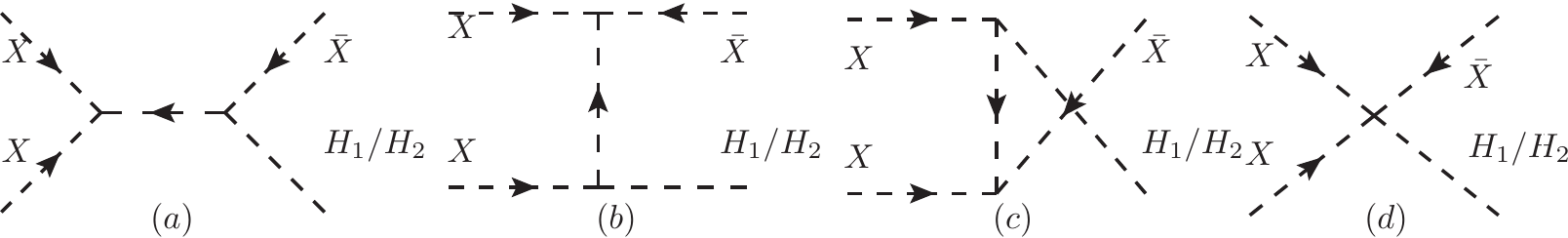}
\includegraphics[width=0.66\textwidth]{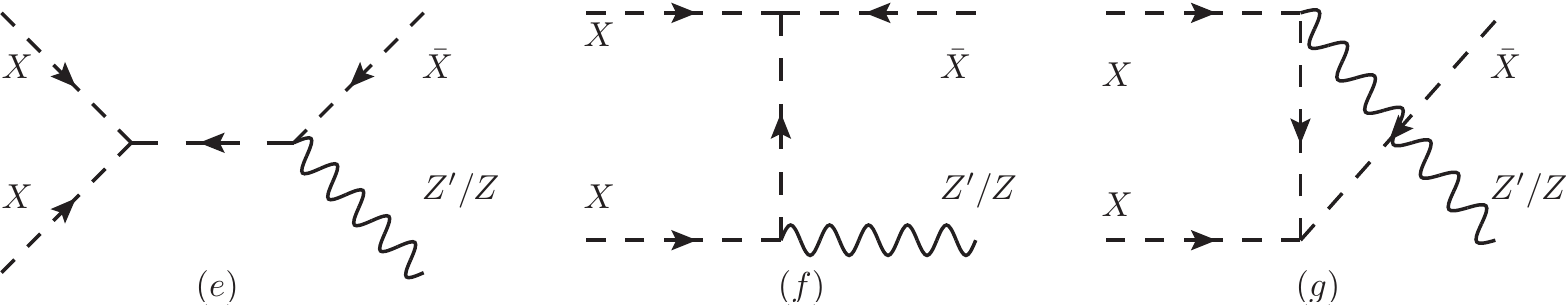} 
\caption{Feynman diagrams for dark matter semi-annihilation. Only (a), (b), and (c) with $H_1$ 
as final state appear in the global $Z_3$ model, while all diagrams could contribute in local $Z_3$ model.
\label{fig:semi-annihilation}}
\end{figure}

\begin{figure}
\includegraphics[width=0.40\textwidth]{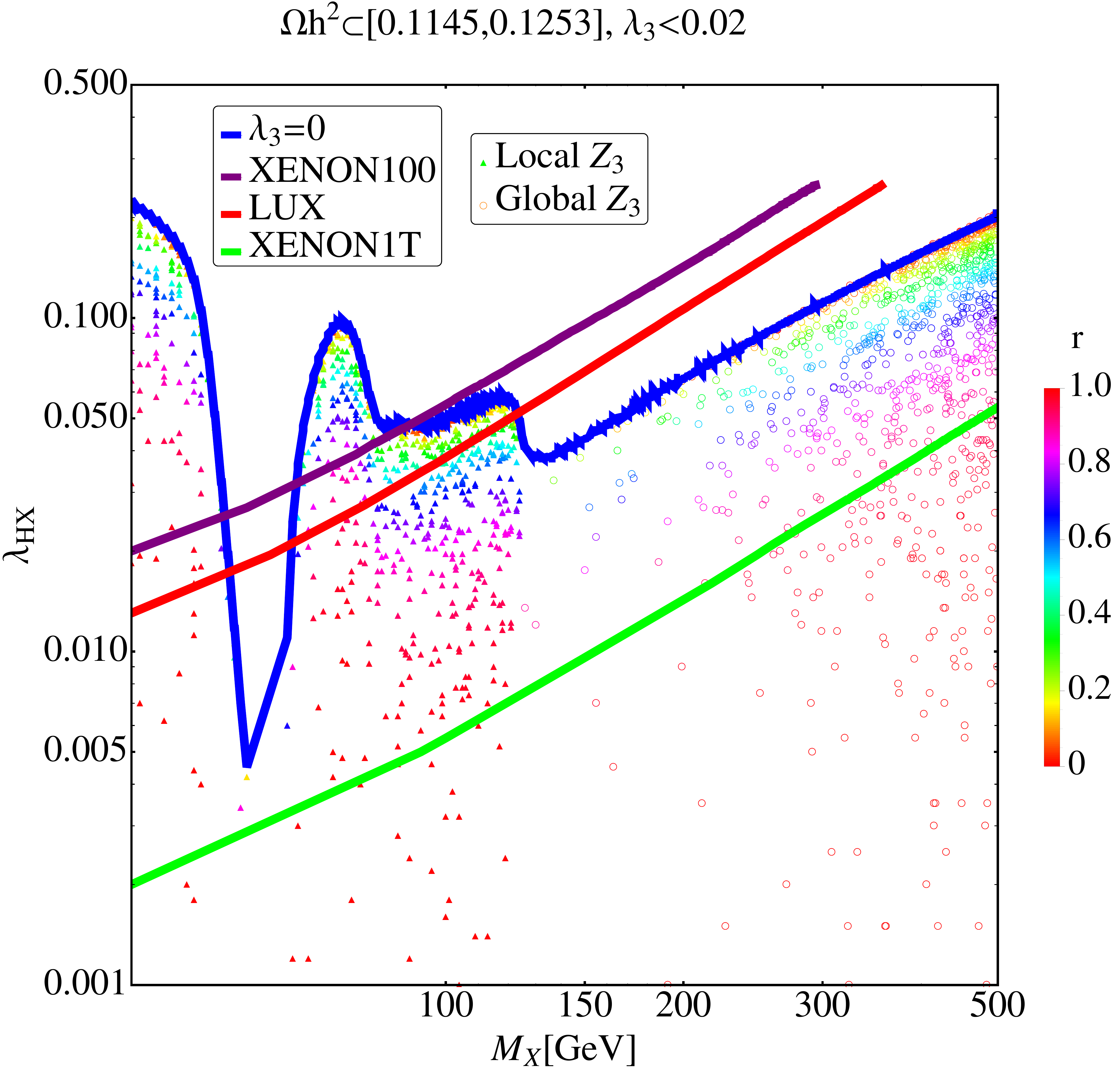}
\caption{Illustration of discrimination between global and local $Z_3$ symmetry. We have chosen 
$M_{H_2}=20$GeV, $M_{Z'}=1$TeV and $\lambda_3<0.02$ as an example. 
Colors in the scatterred triangles and circles indicate the relative contribution of semi-annihilation, 
$r$ defined in Eq.~(9).  The curved blue band, together with the cirles, gives correct relic density 
of $X$ in the global $Z_3$ model. And the colored triangles appears only in the local $Z_3$ model. 
\label{fig:global_gauge}}
\end{figure}

In Fig.~\ref{fig:semi-annihilation}, I show the Feynman diagrams relevant for thermal relic density 
of local $Z_3$  DM $X$. If we worked in global $Z_3$ DM model instead, we would have diagrams 
only with $H_1$ in (1),(b) and (c).  For local $Z_3$ model, there are two more new fields, dark Higgs 
$H_2$ and dark photon $Z^{'}$, which can make the phenomenology of local $Z_3$ case completely 
difference from that of global $Z_3$ case.  In fact, this can be observed immediately in 
Fig.~\ref{fig:global_gauge}, where the open circles are allowed points in global $Z_3$ model, whereas
the triangles are allowed in local $Z_3$ case. The main difference is that in global $Z_3$ case, the same 
Higgs portal coupling $\lambda_{HX}$ enters both thermal relic density and direct detections.  And 
the stringent constraint from direct detection forbids the region for DM below 120 GeV.  On the other hand
this no longer true in local $Z_3$ case, and there are more options to satisfy all the constraints
~\cite{Ko:2014nha,Ko:2014loa}.

We may define the fraction of the contribution from the semi-annihilation in terms of  
\begin{equation}
\label{eq:r} 
r\equiv\frac{1}{2}\frac{v\sigma^{XX\rightarrow X^{\ast}Y}}{v\sigma^{XX^{\ast}\rightarrow YY}+\frac{1}{2}v\sigma^{XX\rightarrow X^{\ast}Y}}.
\end{equation}

Also one can drive the low energy EFT and discuss its limitation, the details
of which can be found in Ref.~\cite{Ko:2014nha}.  The main message is that the EFT cannot enjoy 
the advantages of having the full particles spectra in the gauge theories, namely not-so-heavy dark Higgs
and dark gauge bosons, which could be otherwise helpful for explaining the galactic center 
$\gamma$-ray excess or the strong self-interacting DM.  
And it is important to know what symmetry stabilizes the DM particles.

\begin{table}
\begin{tabular}{|c||c|c|}
\hline
           & Global $Z_3$ & Local $Z_3$   \\ \hline  \hline
Extra fields &  $X$ & $X, Z^{'}, \phi$    \\   \hline
Mediators & $H$ & $H$, $Z^{'}$, $\phi$    \\   \hline
Constraints & Direct detection   & Can be relaxed \\
                   & Vacuum stability &  Can be relaxed \\    \hline
DM mass & $m_X \gtrsim 120$GeV  & $m_X < m_H$ allowed     
\\ 
\hline  
\end{tabular}
\caption{\label{tab:z3} 
Comparison between the global and the local $Z_3$ scalar dark matter models.
Here $X$ is a complex scalar DM,  $H$ is the observed SM-HIggs like boson, and $\phi$ is the dark 
Higgs from $U(1)_X$  breaking into $Z_3$ subgroup. 
}
\end{table}


\subsection{Other possibilities}

Sterile neutrinos including the RH neutrinos are natural candidates for hidden sector fermions 
with dark gauge charges.
In fact there have been some attempt to construct models for CDM interacting with sterile 
neutrinos in order to solve the some puzzles in the standard CDM paradigm as well as 
to reconcile the amount of dark radiation reported by Planck observation and the sterile 
neutrino masses and mixings that fit the neutrino oscillation data~\cite{Ko:2014bka}. 
One can also consider unbroken  $U(1)_X$ dark gauge symmetry with scalar DM and 
the RH neutrinos decay both to the SM and the dark  sector particles~\cite{Baek:2013qwa}.  

\section{Stable DM due to topology: Hidden sector monopole and vector DM, dark radiation}

In field theory there could be a topologically stable classical configuration.  The most renowned 
example is the 't Hooft-Polyakov monopole. This object in fact puts a serious problem in cosmology,
and was one of the motivations for inflationary paradigm. 
In Ref.~\cite{Baek:2013dwa}, we revived this noble idea by putting the monopole in the hidden 
sector and introducing the Higgs portal interaction to connect the hidden and the visible sectors.

Let us consider $SO(3)_X$-triplet real scalar field $\vec{\Phi}$ with the following Lagrangian implemented 
to the SM: 
\begin{equation} \label{Lag}
{\cal L}_{\rm new}  = - \frac{1}{4} V_{\mu\nu}^a V^{a \mu\nu} + \frac{1}{2} 
D_\mu \vec{\Phi} \cdot D^\mu \vec{\Phi}    -  \frac{\lambda_\Phi}{4} 
\left( \vec{\Phi} \cdot \vec{\Phi} - v_\phi^2 \right)^2 
 - \frac{\lambda_{\Phi H}}{2}   \left(\vec{\Phi} \cdot \vec{\Phi} - v_\phi^2\right)
\left( H^\dagger H - \frac{v_H^2}{2} \right) .
\end{equation}
The Higgs portal interaction is described by the $\lambda_{\Phi H}$ term, which is a new addition to the 
renowned 't Hooft-Polyakov monopole model.  

After the spontaneous symmetry breaking of $SO(3)_X$ into $SO(2)_X (\approx U(1)_X)$ by nonzero 
vacuum expectation value (VEV) of $\vec{\Phi}$ with  $\langle \vec{\Phi} (x) \rangle = ( 0 , 0 , v_\Phi )$, 
hidden sector particles are composed of massive dark vector bosons $V_\mu^\pm$ 
\footnote{Here $\pm 1$ in $V_\mu^\pm$ indicate the dark charge 
under $U(1)_X$, and not ordinary electric charges.} with masses $m_V=g_X v_\Phi$ (which are stable 
due to the unbroken subgroup $SO(2)_X \approx U(1)_X$),  massless dark photon 
$\gamma_{h, \mu} \equiv V_\mu^3$,  topologically stable heavy (anti-)monopole  with mass 
$m_M \sim m_V/\alpha_X$, and massive real scalar  $\phi$ (dark Higgs boson) mixed with the SM 
Higgs boson through the Higgs portal term. 

Note that there is no kinetic mixing between $\gamma_h$ and the SM {$U(1)_Y$-gauge boson unlike  
the $U(1)_X$-only case,   due to the non Abelian nature of the hidden gauge symmetry.  Also the VDM is 
stable even in the presence of nonrenormalizable operators due to the unbroken subgroup $U(1)_X$. 
This would not have been the case, if the $SU(2)_X$ were completely broken by a complex 
$SU(2)_X$ doublet,  where the stability of massive VDM is not protected by $SU(2)_X$ gauge symmetry 
and nonrenormalizable  interactions would make the VDM  decay in general~\cite{Hambye:2008bq}.  
Of course, it would be fine 
as long as the lifetime of the decaying VDM is long enough so that it can  still be a good CDM candidate.
In the VDM model with a hidden sector monopole, the unbroken $U(1)_X$ subgroup not only protects 
the stability of VDM $V_\mu^\pm$, but also contributes to the dark radiation at the level of $\sim 0.1$.  
We refer the readers to the original paper on more details of phenomenology of this model
~\cite{Baek:2013dwa}. 

\section{EWSB and CDM from Strongly Interacting Hidden Sector:   \\
long-lived DM due to accidental symmetries} 

Another nicety of models with hidden sector is that one can construct a model 
where all the mass scales of the SM particles and DM are generated by 
dimensional transmutation in the hidden sector~\cite{Hur:2007uz,Ko:2008ug,Hur:2011sv}. 
Basically the light hadron masses such as proton or $\rho$ meson come from confinement, which is 
derived from massless QCD through dimensional transmutation.  One can ask if all the masses of 
observed particles can be generated by quantum mechanics, in a similar manner with the proton mass
in the massless QCD.  The most common way to address this question is to employ the Coleman-Weinberg
mechanism for radiative symmetry breaking.  Here I present a new model based on nonperturbative 
dynamics  like technicolor or chiral symmetry breaking in ordinary  QCD.

Let us consider a scale-invariant extension of the SM with a strongly interacting hidden sector: 
\begin{eqnarray}
  {\cal L} &  = & {\cal L}_{\rm SM, kin} + {\cal L}_{\rm SM, Yukawa}
- {\lambda_{H} \over 4} ( H^{\dagger} H )^2  - {\lambda_{SH} \over 2}~S^2 H^{\dagger} H 
- {\lambda_S \over 4} S^4  \nonumber \\ 
& - & \frac{1}{4} {\cal G}_{\mu\nu}^a {\cal G}^{a \mu\nu}  
+ \sum_{k=1,...,f} \overline{\cal Q}_k \left[   i D \cdot \gamma  - \lambda_k S 
\right] {\cal Q}_k .
\end{eqnarray}
Here ${\cal Q}_k$ and ${\cal G}_{\mu\nu}^a$ are the hidden sector quarks and gluons, and 
and the index $k$ is the flavor index in the hidden sector QCD. 
In this model, we have assumed that the hidden sector strong interaction is vectorlike and confining 
like the ordinary QCD.  Then we can use the known aspects of QCD dynamics to the hidden sector QCD.

Note that the real singlet scalar $S$  plays the role of messenger connecting 
the SM Higgs sector and the  hidden sector quarks.  

In this model, dimensional transmutation in the hidden sector will generate the
hidden QCD scale and chiral symmetry breaking with developing
nonzero $\langle \bar{\cal Q}_k {\cal Q}_k \rangle$. 
 Once  a nonzero $\langle \bar{\cal Q}_k {\cal Q}_k \rangle$ is developed, 
the  $\lambda_k S $ term generate the linear potential for the real  singlet $S$, 
leading to nonzero $\langle S \rangle$. This in turn generates the hidden sector current 
quark masses through $\lambda_k$ terms as well as the EWSB through $\lambda_{SH}$ term.
Then the Nambu-Goldstone boson $\pi_h$ will get nonzero masses, and becomes 
a good CDM  candidate. Also hidden sector baryons ${\cal B}_h$ will be formed, 
the lightest of which would be long lived due to the accidental h-baryon conservation.
See Ref.~\cite{Hur:2011sv} for more details.

\section{Light mediators and Self-interacting DM}

Another nice feature of the dark matter models with local dark gauge symmetry is that the model 
includes new degrees of freedom,  dark gauge bosons  and dark Higgs boson(s), that can play 
the role of force mediators from the beginning because of the rigid structure of the underlying gauge 
theories.  
In fact one can utilize the light mediators in order to explain the GeV scale $\gamma$-ray excess 
or the self-interacting DM which would solve three puzzles in the CDM paradigm: 
(i) core-cusp problem, (ii) missing satellite problem and (iii) too-big-to-fail problem.
These would have been simply impossible if we adopted the EFT approach for DM physics.

In the EFT approach for the DM, these new degrees of freedom are very heavy compared with the 
DM mass as well as the energy scale we are probing the dark sector (e.g., the collider energy scale).  
However, we don't know anything about the mass scales of these mediators, and it would be too strong 
an assumption. Without these light mediators, we could not explain the GeV scale $\gamma$-ray excess
as described in this talk, or have strong self-interacting DM.  This illustrates one of the limitations of DM 
EFT appraoches. 

\section{Higgs inflation assisted by the Higgs portal }

The final issue related with DM models with local dark gauge symmetris is the Higgs inflation 
in the presence of the Higgs portal interaction to the dark sector:   
\begin{equation} \label{L-jordan}
\frac{\mathcal{L}}{\sqrt{-g}} = - \frac{1}{2 \kappa} \left( 1 + \xi \frac{h^2}{M_{\rm Pl}^2} 
\right) R + \mathcal{L}_h  + \lambda_{\phi H} \phi^2 h^2  
\end{equation}
in the unitary gauge, 
where $\kappa = 8 \pi G = 1/M_{\rm Pl}^2$ with $M_{\rm Pl}$ being the reduced Planck mass, 
and $\mathcal{L}_h$ is the Lagrangian of the SM Higgs field only.    Here $\phi$ denotes a generic 
dark Higgs field which mixes with the SM Higgs field after dark and EW gauge symmetry breaking. 
In the presence of the Higgs portal interaction, we have recalculated the slow-roll parameters.
Relegating the details to Ref.~\cite{Ko:2014eia},  I simply show the results: 
at a bench mark point for Fig.~2 of Ref.~\cite{Ko:2014eia}, 
we get the following results: 
\begin{equation}
n_s = 0.9647 \ , \ r = 0.0840 \ , \  
\end{equation}
for $N_e = 56$, $h_* / M_{\rm Pl} = 0.72$, $\alpha =  0.07422199$ and $\xi = 12.8294$ 
for a pivot scale $k_*=0.05 {\rm Mpc}^{-1}$.
There is a parameter space where the spectral running of $n_s$ is small enough at the level of
$|  n_s^{'} | \lesssim 0.01$.  It is amusing to notice that the $r$ could be as large as 
$\sim O(0.1)$ in the presence of the Higgs portal interactions to a dark sector, independent of 
the top quark and the Higgs boson mass in the standard Higgs inflation scenario.




\section{Higgs phenomenology, EW vacuum stability,  and dark radiation}
Now let us discuss Higgs phenomenology within this class of DM models. 
Due to the mixing effect between the dark Higgs and the SM Higgs bosons, 
the signal strengths of the observed Higgs boson will be universally reduced from ''1''  
independent of production and decay channels~\cite{Baek:2011aa,Baek:2012se}. 
Also the 125 GeV Higgs boson could decay into a pair of dark Higgs and/or a pair of dark gauge boson, 
which is still  allowed by the current LHC data~\cite{Chpoi:2013wga}. These predictions 
will be further constrained by the next round experiments. 

Also the dark Higgs can make the EW vacuum stable upto the Planck scale without 
any other new physics~\cite{Baek:2012uj,Baek:2012se}, and this was very important in the 
Higgs-portal assisted Higgs inflation discussed in the previous  section.

In most cases, there is generically a singlet scalar which is nothing but a dark Higgs, which would give 
a new motivation to consider singlet extensions of the SM. Traditionally a singlet scalar was 
motivated mainly  by why-not or $\Delta \rho$ constraint, or the strong first order EW phase 
transition for electroweak baryogenesis.
Being a singlet scalar, the dark Higgs will satisfy all these motivations, as well as stability of DM 
by local dark gauge symmetry.   
It would be important to seek for this singlet-like scalar at the LHC or the ILC, but the colliders 
cannot cover the entire mixing angle down to $\alpha \sim 10^{-8}$ (for MeV dark Higgs) relevant to 
DM phenomenology. 

Massless dark gauge boson or light dark fermions in hidden sectors could 
contribute to dark radiation of the universe  
In a class of models we constructed, the amount of extra dark radiation is rather small
by an amount consistent with the Planck data due to Higgs portal  interactions
~\cite{Baek:2013qwa,Baek:2013dwa,Ko:2014bka}.

\section{Collider Search for Dark Higgs: Beyond the DM EFT and simplified models}

Finally let us discuss the collider search for the dark Higgs boson and DM particles.
A classic signature for DM search would be mono $X$ +  missing $E_T$.  
Early this year ATLAS and CMS reported such studies in the monojet + missing $E_T$ 
and $t\bar{t}$ + missing $E_T$, respectively.  Their analyses are based on the simplified
model without the full SM gauge symmetry, which is neither renormalizable nor unitary. 

Let us consider a scalar $\times$ scalar operator describing the direct detection of DM on nucleon,  assuming the DM is a Dirac fermion $\chi$ with some conserved quantum 
number stabilizing $\chi$: 
\begin{equation}
{\cal L}_{SS} \equiv \frac{1}{\Lambda_{dd}^2} \bar{q} q \bar{\chi} \chi 
~~~{\rm or}~~~ \frac{m_q}{\Lambda_{dd}^3} \bar{q} q \bar{\chi} \chi.
\label{eq:operator}
\end{equation}
Assuming the complementarity among direct detection, collider search and indirect detection (or thermal 
relic density),  the bound on the scale $\Lambda_{dd}$ of this operator has been studied extensively 
in literature~\cite{Baek:2015lna}. 

However, the above operator does not respect the full SM gauge symmetry and thus is not suitable for studying phenomenology at high energy scale (say, at electroweak scale).   
Therefore the operator form has to be mended.  
Note that the SM quark bilinear part in the above operator can be written into 
$\overline{Q}_L H d_R ~~~{\rm or} ~~~\overline{Q}_L \widetilde{H} u_R$, 
if we impose the full SM gauge symmetry.  Here $Q_L \equiv ( u_L , d_L  )^T$. 
Likewise, the singlet fermion $\chi$ cannot have renormalizable 
couplings to the SM Higgs boson, since $\chi$ is a singlet whereas
the Higgs field is a doublet. Similarly, the quark bilinear $\bar{q}q$
does not have renormalizable couplings to a singlet scalar field $S$.

The simplest way to write down a renormalizable operator that is invariant 
under the full SM gauge group is to introduce a real signet scalar 
field $S$~~\cite{Baek:2011aa,Baek:2012se}
and induce an operator
$s \bar{\chi} \chi \times h \bar{q} q \rightarrow  \frac{1}{m_s^2} \bar{\chi} \chi \bar{q} q $ 
by integrating out the real scalar $s$.
However there is always a mixing between the SM Higgs $h$ and the real singlet scalar $s$, which results in two physical neutral scalars
$H_1$ and $H_2$ with the mixing angle $\alpha$. 
Therefore one should take into account the exchange of both $H_1$ and $H_2$
for DM direct detection scattering~\cite{Baek:2011aa}.
Note that there is a generic cancellation between two contributions from
two neutral scalars, which cannot be seen 
within EFT approach~\cite{Baek:2011aa,Baek:2012se}.

Let us consider the Higgs portal fermion DM model as an example.
The simplest UV completion is given by Eq.~(\ref{eq:Lag2}), and one can calculate 
the $\psi q \rightarrow \psi q$ scattering amplitude therein:
The interaction Lagrangian of $H_1$ and $H_2$ with the SM fields and DM $\chi$ is 
given by 
\begin{equation} \label{eq:portal2}
{\cal L}_{\rm int} = - ( H_1 \cos\alpha + H_2 \sin\alpha ) \left[ \sum_f \frac{m_f}{v_H} 
 \bar{f} f  - \frac{2 m_W^2}{v_H} W_\mu^+ W^{-\mu} - \frac{m_Z^2}{ v_H} Z_\mu Z^\mu   
 \right] + \lambda ( H_1 \sin\alpha - H_2 \cos\alpha ) \bar{\chi}\chi \ ,
 \end{equation}
following the convention of Ref.~\cite{Baek:2011aa}.  We identify the observed 125 GeV
scalar boson as $H_1$.  The mixing between $h$ and $s$ leads to the universal 
suppression of the Higgs signal strengths at the LHC, 
independent of production and decay channels ~\cite{Baek:2011aa}.

Let us start with the DM-nucleon scattering amplitude at parton level, 
$\chi (p) + q (k) \rightarrow \chi (p') + q (k')$, the parton level amplitude of which is given by 
\begin{eqnarray}
{\cal M} & = & - \overline{u(p')} u(p) \overline{u(k')} u(k) ~\frac{m_q}{v_H} \lambda \sin\alpha \cos\alpha ~
\left[  \frac{1}{t - m_{H_1}^2 + i m_{H_1} \Gamma_{H_1}} - \frac{1}{t - m_{H_2}^2 + i m_{H_2} \Gamma_{H_2}}
\right] \label{eq:tchannel}   \\
&  \rightarrow &  \overline{u(p')} u(p) \overline{u(k')} u(k)  ~\frac{m_q}{2 v_H}  \lambda \sin 2\alpha 
\left[  \frac{1}{m_{H_1}^2}  - \frac{1}{m_{H_2}^2}    \right] 
\equiv \frac{m_q}{\Lambda_{dd}^3} \overline{u(p')} u(p) \overline{u(k')} u(k),
\label{eq:relation}
\end{eqnarray}
where $t \equiv (p' - p)^2$ is  the square of the 4-momentum transfer 
to the nucleon, and we took the limit
$t\rightarrow 0$ in the second line, which is a good approximation to the DM-nucleon scattering.
The scale of the dim-7 effective operator, $m_q\, \bar{q} q\, \overline{\chi}\chi$, describing the direct detection cross section for the DM-nucleon scattering is defined in terms of 
$\Lambda_{dd}$:
\begin{eqnarray}
\Lambda_{dd}^3 & \equiv & \frac{2 m_{H_1}^2 v_H}{ \lambda  \sin 2\alpha} \left(  1 - \frac{m_{H_1}^2}{m_{H_2}^2}  
\right)^{-1},  \\
\bar{\Lambda}_{dd}^3 & \equiv & \frac{2 m_{H_1}^2 v_H}{ \lambda  \sin 2\alpha},
\label{eq:BarlambdaDD}
\end{eqnarray}
where $\bar{\Lambda}_{dd}$ is derived from $\Lambda_{dd}$ in the limit 
$m_{H_2} \gg  m_{H_1}$.
It is important to notice that the amplitude~(\ref{eq:tchannel}) 
was derived from renormalizable and unitary 
Lagrangian with the full SM gauge symmetry, and thus can be a good starting point for addressing
the issue of validity of complementarity. 

The amplitude for the monojet with missing transverse energy($\met$) signature at hadron colliders is 
connected to the amplitude~(\ref{eq:tchannel}) by crossing symmetry  $s\leftrightarrow t$.  
Comparing with the corresponding  amplitude from the EFT approach, we have to include 
the following form factor:  
\begin{equation}
\frac{1}{\Lambda^3_{dd}} \rightarrow \frac{1}{\bar{\Lambda}^3_{dd}} \left[ 
\frac{m_{H_1}^2}{ \hat{s} - m_{H_1}^2 + i m_{H_1} \Gamma_{H_1}}  - 
\frac{m_{H_1}^2}{ \hat{s} - m_{H_2}^2 + i m_{H_2} \Gamma_{H_2}} \right]  
\equiv  \frac{1}{\Lambda^3_{col} ( \hat{s} )}  ,
\label{eq:collisionS}
\end{equation}
where $\hat{s} \equiv m_{\chi\chi}^2$ is  the square of the invariant mass  of the DM pair.
Note that $s \geq 4 m_\chi^2$ in the physical 
region for DM pair creation,  and that there is no single constant scale  
$\Lambda_{col}$  for an effective operator that characterizes the 
$q\bar{q} \rightarrow \chi \bar{\chi}$, since $\hat{s}$ varies in the range of 
$4 m_\chi^2 \leq \hat{s} \leq s$ with $\sqrt{s}$ being the center-of-mass (CM) energy 
of the collider.   
Also note that we have to include two scalar propagators with opposite sign in order to 
respect the full SM gauge symmetry  and renormalizability. This is in sharp contrast with 
other  previous studies where only a single propagator is introduced to replace 
$1/\Lambda^2$.    
The two propagators interfere destructively for very high $\hat{s}$
or small $t$ (direct detection), but for $m_{H_1}^2 < \hat{s} <m_{H_2}^2$,
they interfere constructively.
The $1/s$ suppressions from the $s$-channel resonance 
propagators  make the amplitude unitary, in compliance with renormalizable and unitary QFT.

If one can fix $\hat{s}$ and $m_{H_2}^2 \gg \hat{s}$, we can ignore  the  2nd propagator. 
But at hadron colliders, $\hat{s}$ is not fixed, 
except for the kinematic condition $4 m_\chi^2 \leq \hat{s} \leq s$ 
(with $s = 14$TeV for example at the LHC@14TeV).   
Therefore we cannot say clearly when we can ignore $\hat{s}$ compared with 
$m_{H_2}^2$ at hadron colliders, unless $m_{H_2}^2 > s$ (not $\hat{s}$).  

One can derive the bound on the effective mass scale $M_*$ within the full renormalizable
and unitary models and compared with the bounds derived with the EFT approaches, 
with the same $\overline{\Lambda}_{dd}$.  The results are shown in Fig.~3: 
the left panel on  the monojet + $\ET$ from ATLAS data and the right panel on the 
$t\bar{t} + \ET$ from the CMS data. The blue lines are the results from the simplified model
with a singlet scalar propagator, and the red lines are those from the renormalizable and 
unitary (and gauge invariant for the VDM) models. Note that the bounds depend very much 
on the  underlynig model assumption, and are sensitive to the 2nd scalar boson, which does
not appear in the EFTW or the usual simplified model.  
These plots show that it is very important to analyze the monojet + $\ET$ and $t\bar{t} + \ET$ 
data from the LHC within well-defined renormalizable, unitary and gauge invariant DM models.
The usual EFT and the simplified models without the full SM gauge symmetry do not describe
DM physics at high energy colliders properly.

 \begin{figure}[th]
\includegraphics[width=0.50\textwidth]{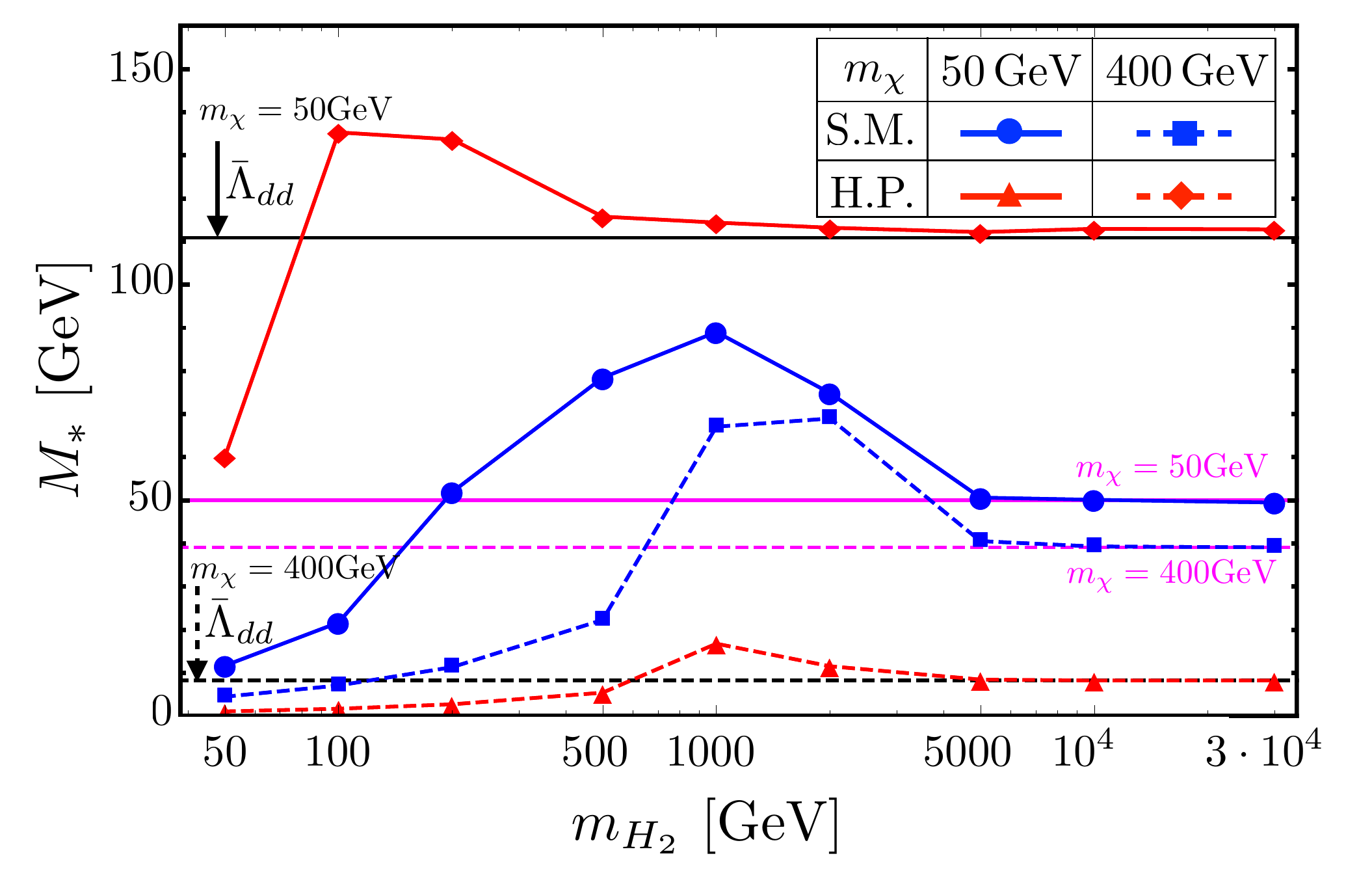}
\includegraphics[width=0.50\textwidth]{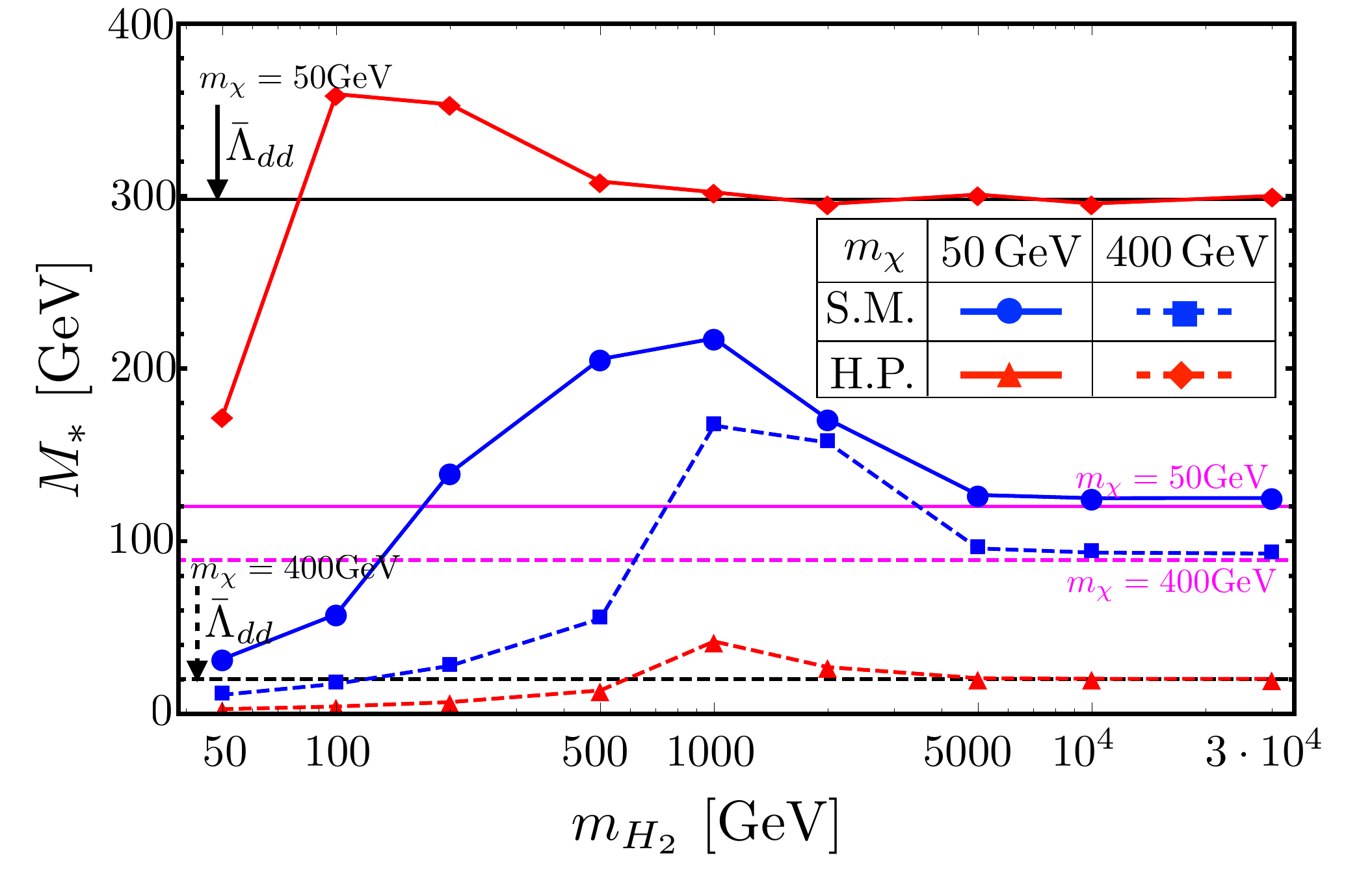}
\caption{Observed exclusion limits in terms of $m_{\chi}$ and $M_*$ 
with $90\%$ CL. from mono-jet+$\met$ search (left) 
and $t\bar t+\met$ search (right). }
\label{fig:simp}
\end{figure}
\section{Conclusion and Outlook}

In this talk, I discussed a class of dark matter models where dark gauge symmetry plays
an important role in stabilizing electroweak scalar DM or making them long lived enough 
compared with the age of the universe. I discussed three explicit examples: (i) DM is stable
due to unbroken dark gauge symmetry  $Z_3$ originating from $U(1)_X$ gauge symmetry, 
(ii) DM is stable due to topological reason, the famous 't Hooft-Polyakov monopole in the
hidden sector, and the unbroken $U(1)$ subgroup gaurantees the stability of the vector DM
in the monopole sector, and (iii) DM is long lived due to global flavor symmetry which is 
an accidental symmetry of underlying new strong interaction in the dark sector.
I also discussed the limitation of the DM EFT or simplified DM models, which are not either
renormalizable or not invariant under the full SM gauge invariance. Both of them are 
important in the DM model building for studying DM phenomenology at high energy colliders.
Also dark Higgs or dark gauge bosons can play important role in DM self-interaction or 
galactic center $\gamma$-way execss, which are not possible in the 

One of the generic predictions of the Higgs portal DM models and hidden sector DM models 
with local dark gauge symmetry is the existence of a new neutral scalar boson which is mostly 
the SM singlet if the DM particles are either fermion or vector.  
It affects the DM signatures at high energy  colliders because of the form factors with two scalar 
propagators with negative sign,   Eq. (21).  This feature is a consequence of the full SM gauge 
invariance and renormalizability,  and can not be seen in the usual EFT approach or simplified 
DM models.  The detailed study of the Higgs portal DM phenomenology at future colliders 
will be presented elsewhere.  


\acknowledgments
The author is grateful to Takayama  for the kind invitation to DSU2015.
He also thanks Seungwon Baek, Taeil Hur, Dong Won Jung, Myeonghun Park, 
Wan Il Park,  Eibun Senaha and Yong Tang for enjoyable collaborations on the subjects presented 
in this talk.   This work is supported in part by National Research Foundation of Korea (NRF) 
Research Grant NRF-2015R1A2A1A05001869


\end{document}